\documentclass[osajnl,twocolumn,showpacs,groupedaddress,10pt]{revtex4-1}
\usepackage{graphicx}

\begin{document}
\title{Tunable Rydberg excitons maser}
\author{D. Ziemkiewicz}
\email{Corresponding author: david.ziemkiewicz@utp.edu.pl}
\author{S. Zieli\'{n}ska-Raczy\'{n}ska}
\affiliation{Institute of Mathematics and Physics, UTP University of Science and Technology, \\ Al. Prof. S. Kaliskiego 7, 85-789
Bydgoszcz, Poland}




\begin{abstract}
We propose a solid state maser based on Cu$_2$O, where ensemble of highly excited Rydberg exciton states serves as a gain medium. We show that the system is highly tunable with external electric field, allowing for a wide range of emission frequencies. Numerical simulations of system dynamics are performed to optimize the conditions for efficient masing and estimate the emission power.
\end{abstract}


\maketitle

\section{Introduction}
Excitons, bound pairs of electrons and holes, are the basic excitations of semiconductors.

Rydberg excitons (RE) in cuprous oxide, discovered in 2014 by Kazimierczuk et al \cite{Kazimierczuk}, are highly excited states with many unique properties. The exciton Rydberg energy of about 90 meV is lower by two order of magnitude as compared to the atomic Rydberg energy and this reduction makes excitons sensible to external fields. RE offer a promising combination of huge size, much greater then wavelength of light needed to create them, with long lifetimes which scale as $n^2$ ($\tau = 28$ ns for n=20), wide spectrum of wavelength corresponding  to transitions between their eigenstates and  the energy spacing of neighbouring states, which decreases as $n^{-3}$. As in all Rydberg systems, the occurrence of RE within smaller distance is prevented by the exciton Rydberg blockade which arises from the dipole-dipole interactions between them. Due to this interaction, if an exciton is created, the energy of exciting another exciton in the vicinity is shifted far out of the resonance.


The majority of research involving in Rydberg excitons, both theoretically and experimentally \cite{Kazimierczuk}, has been concentrated on their static, spectroscopic properties (such as resonaces \cite{My1}, splitting and crossing of RE states in external fields \cite{My2}).   Recently, a single photon source based on RE in cuprous oxide has been proposed \cite{Khazali} and also RE in semiconductor microcavities were investigated in context of achieving giant optical nonlinearities \cite{Walther}.

Due to the fact that high Rydberg excitons can interact resonantly and very strongly with milimeter-wave radiation, we aim to investigate the dynamics of such a medium in the situation of population inversion, which leads to possible realization of a solid state analogue of Rydberg atom maser \cite{Moi}. In this paper we propose theoretically a maser based on Cu$_2$O crystal with highly excited Rydberg excitons as active medium.

Recent experimental verifcation of solid state maser based on diamond \cite{Breeze}\cite{Jin}, has awaked  a revival of interest of this topic. While in \cite{Breeze} the authors have demonstrated continuous maser operation at room temperature, we propose to use ensembles of optically pumped RE as a gain medium  which operates at low temperature (about several K), but due to the several unique characteristics of RE medium, namely exceptionally long lifetimes of excitons in their highly excited states and their strong coupling to external field, RE in Cu$_2$O are promising candidates for realization of a highly tunable, moderate power device. The wealth of accessible states provides many transitions within millimetre range wavelength while high density of excitons might lead to a significant output power and opens new possibilities in the domain of sub-mm wave amplification and detection.

We recall a simple semiclassical theory of maser operation with respect to the specific case of RE in Cu$_2$O and optimize the maser parameters discussing the geometry, cavity and choice of excitonic states. Our numerical simulation demonstrates that masing and microwave amplification are feasible in accessible conditions (sub-millimetre sample size, Q$\sim 10^5$, pump power density $10^{-3}$ W/mm$^2$ with the  571 nm optical pump), obtaining up to $10^{-6}$ W emission power in the sub- and millimetre wavelength range.
\section{System model}
\begin{figure}[ht!]
\centering
\includegraphics[width=.7\linewidth]{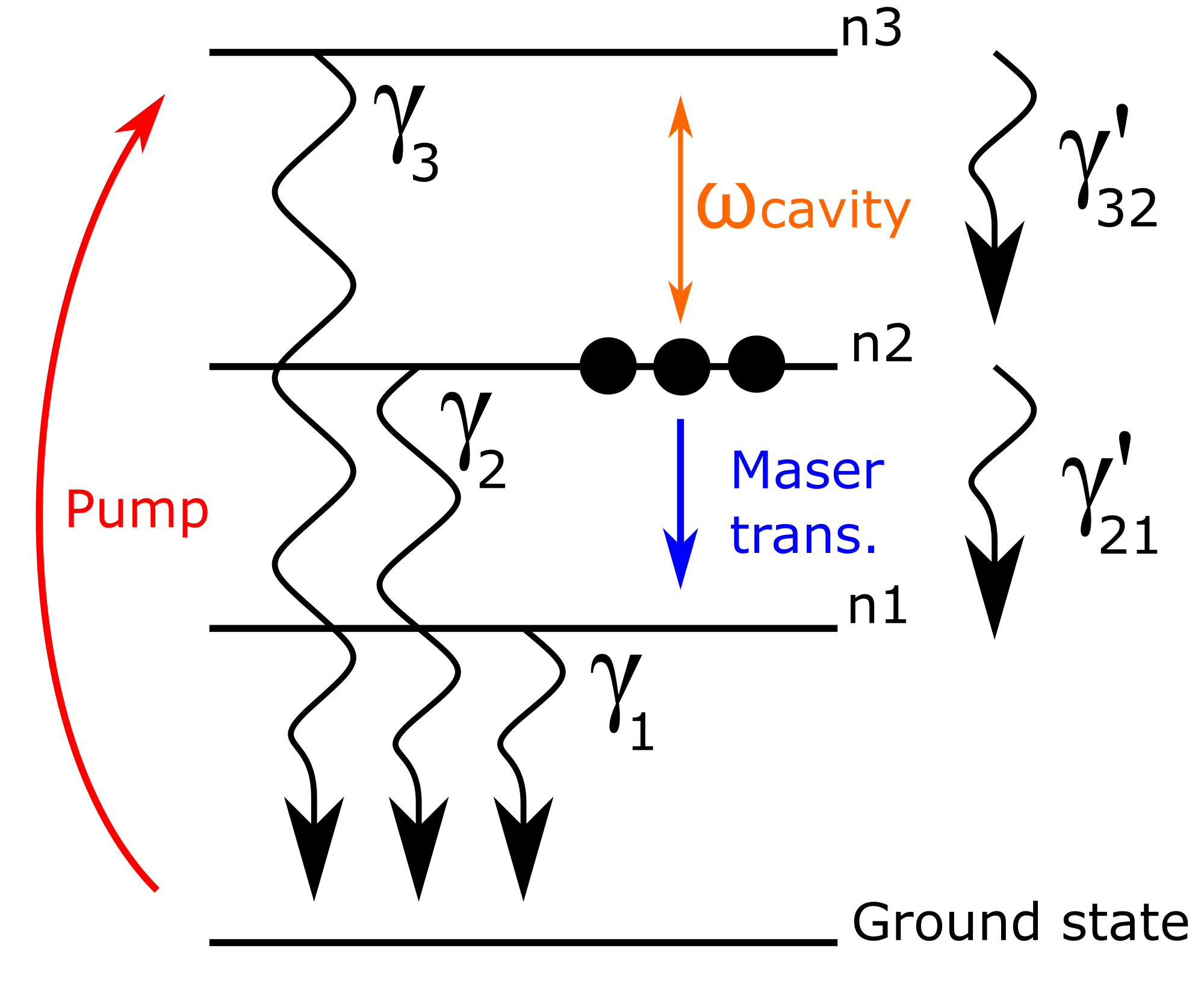}
\caption{Energy levels of the considered system}\label{model}
\end{figure}
Let's consider a Cu$_2$O crystal which placed in a cavity with frequency $\omega_c$ and quality factor $Q$. Out of the accessible excitonic states, we choose three levels $n_1P$, $n_2S$, $n_3P$ with populations $N_1$, $N_2$, $N_3$ accordingly, as shown on the Fig. (\ref{model}). The crystal is subject to external electric field which lifts the degeneracy of $S$ and $P$ states and can be used to tune the energy levels via Stark shift \cite{My2}. In particular, the cavity is tuned to the transition $n_3 \rightarrow n_2$. The EM field energy density is denoted $W$. The population dynamics is described by set of equations
\begin{eqnarray}
\frac{\partial N_3}{\partial t} &=& bP_r - \gamma_3N_3 - \gamma_{32}'N_3\nonumber\\
\frac{\partial N_2}{\partial t} &=& \gamma_{32}'N_3 - \gamma_2N_2 - \gamma_{21}'N_2 - (N_2-N_1)BW\nonumber\\
\frac{\partial N_1}{\partial t} &=& \gamma_{21}'N_2 + (N_2-N_1)BW - \gamma_1N_1\nonumber\\
\frac{\partial W}{\partial t} &=& \hbar\omega(N_2-N_1)BW - 2\gamma_cW \label{r_uklad}
\end{eqnarray}
where
$P_r=P_{pump}/E_{exciton}(n_3)$ is the pump rate,
$\gamma_c=\frac{\omega_c}{2Q}$ is the cavity dissipation rate, $b$ is pump absorption quenching factor which depends on exciton density and drops to 0 then Rydberg blockade is reached. 
$B=\frac{\pi}{3\epsilon_0\hbar^2}|d_{12}|^2P_{12}$ is the Einstein cofficient. This quantity depends on the dipole moment $d_{12}$ and is further modified by the so-called Purcell factor $P_{12}$ \cite{Purcell}. For transition $i \rightarrow j$, of frequency $\omega_{ij}$ and wavelength $\lambda_{ij}$, the Purcell factor is given by \cite{Purcell}
\begin{equation}
P_{ij}=\frac{\lambda_{ij}^3}{V}\frac{3Q}{4\pi^2}\frac{(\omega_{ij}/Q)^2}{4(\omega_{ij}-\omega_c)^2 + (\omega_{ij}/Q)^2}
\end{equation}
where V is the mode volume of the cavity. In the limit of closed system, e. g. $Q >> 1$, it is similar to the geometric volume \cite{Sauvan}. Note that for microwave transitons, $\frac{\lambda_{ij}^3}{V} \sim 1$. Therefore, by applying external electric field, one can use the Stark shift to match the transition frequency to the cavity, e. g. $\omega_{23}-\omega_c \approx 0$ and obtain $P_{23} \sim Q$. The damping rates in Eqs (\ref{r_uklad}) depend on the Einstein's coefficient for spontaneous emission
\begin{equation}
A_{ab} = \frac{4\omega_{ab}^3}{3 \hbar c^3} \frac{max(l_a,l_b)}{2l_a+1}|\langle n_a|er|n_b \rangle|^2P_{12}
\end{equation}
which is also influenced by the factor $P_{12}$. This means that transition rate $\gamma'_{32}$ can be amplified by a factor of $Q \sim 10^5$. This is the key idea of our paper; we take advantage of the fact that the exceptionally long lifetimes of higher Rydberg excitonic states make the $n_2$ level metastable. At the same time, the lifetime of $n_3$ is shortened by Purcell factor, so that the $n_3 \rightarrow n_2$ transition becomes dominant. Therefore, a large population $N_2$ can be maintained even for relatively low pump power, leading to the population inversion. Note that the $n_1$ level is short-living, so that in most cases $N_1 \approx 0$. When a steady state is reached, e. g., $\frac{\partial W}{\partial t}=0$, one can calculate the emission power
\begin{equation}
P = \hbar\omega(N_2-N_1)BW = 2\gamma_cW.
\end{equation} 
It is important to note that due to the Rydberg blockade, there is an upper limit of exciton density, which can be estimated as \hbox{$\rho_{max}=(4/3\pi r_{n}^3)^{-1} \sim n^{-6}$}, where 
\begin{equation}
r_n = \frac{\mbox{1.1 nm}}{2}[3n^2-l(l+1)]
\end{equation}
is the excitonic radius \cite{Kazimierczuk}. This mechanism has been taken into account in factor $b$ which causes exponential drop of pump absorption when the exciton density approaches the critical value $\rho_{max}$.

For the spontaneous emission rate to the ground state, we have used value from data fitting \cite{Kazimierczuk}\cite{My3}
\begin{equation}
\Gamma_n = 240 \mu eV \frac{16(1+0.01n^2)}{n^3}.
\end{equation}
The above values are taken from experimental data and include the effects of scattering on acoustical and optical phonons \cite{Stolz}.

For $n_2 \rightarrow n_1$ transitions, one can use the estimation
\begin{equation}
\Gamma_{12} \approx \Gamma_1 \frac{\omega_{12}^3}{\omega_1^3}.
\end{equation}
since the transition probability scales as $\omega^{3}$ \cite{Gallagher}. The dipole moments $|\langle n_a,l|er|n_b,l\pm 1 \rangle|^2$ have been calculated from the overlap of hydrogen-like wavefunctions with additional quantum defect $\delta = 0.23$ which originates from Cu$_2$O valence band structure \cite{Hoogenraad}\cite{Heck} (see Appendix A). In general, the transition dipole moment is highest for nearby states and scales with the exciton radius \cite{Kazimierczuk}.

\section{Numerical results}
Due to multiple accessible Rydberg excitonic states, there is a significant number of possible combinations of $n_3, n_2, n_1$ states which need to be examined for choosing the optimal conditions for masing. In particular, there are 2300 systems with energy levels $n_1<n_2<n_3<25$.

\begin{figure}[ht!]
\centering
\includegraphics[width=.8\linewidth]{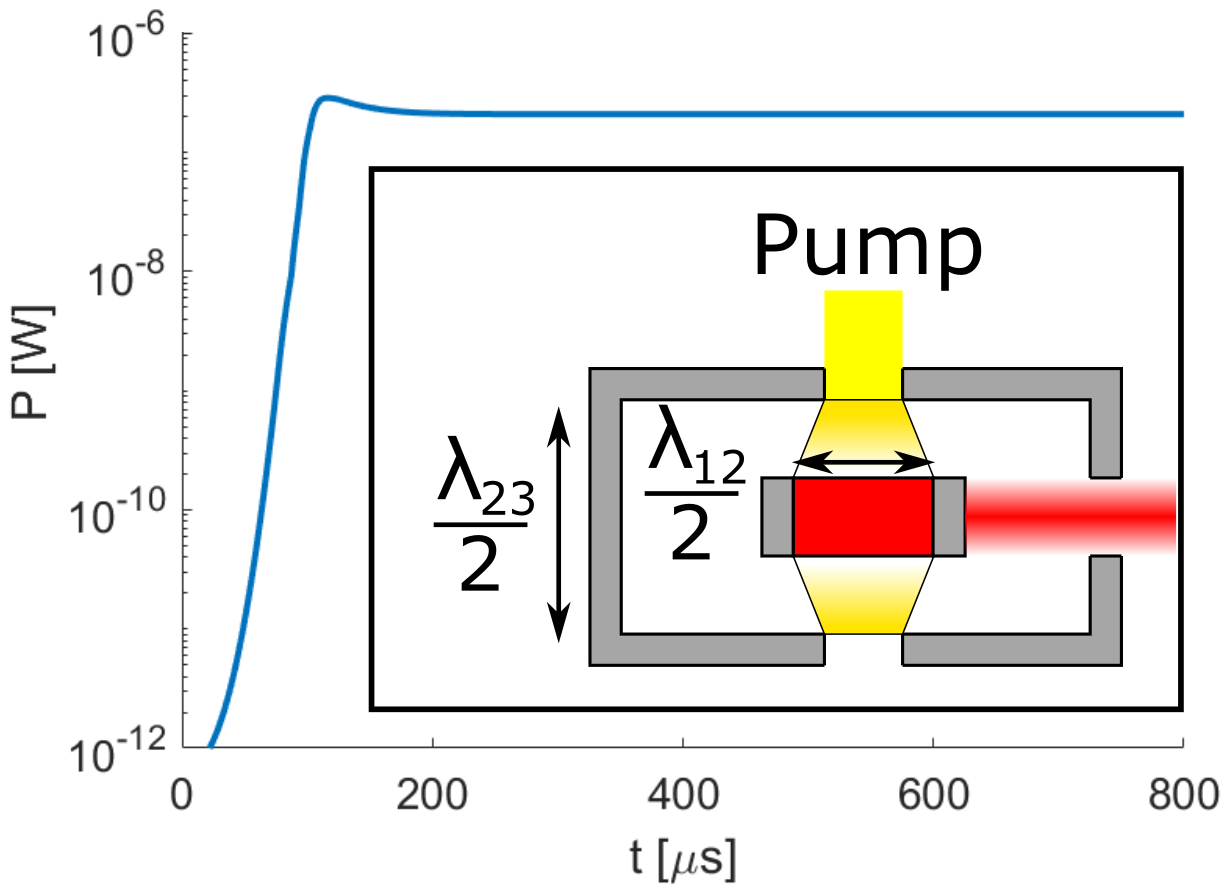}
\caption{Emission power as a function of time. $Q=10^5$, $\rho_{p}=10^{-2}~W/mm^2$. \hbox{$8\rightarrow6\rightarrow5$ states}. Inset: device geometry.}\label{rys_Typ}
\end{figure}

  The Cu$_2$O crystal is placed in a metallic cavity (see Fig. \ref{rys_Typ} inset). The crystal has form of a cylinder with radius $r$=0.2 mm and length $l=n\lambda_{12}/2$, $n \in N$, which is matched to the masing transition $n_2 \rightarrow n_1$, while the cavity is tuned to the $n_3 \rightarrow n_2$ transition, has a volume of $\left(\frac{\lambda_{23}}{2}\right)^3$ and the quality factor $Q=10^5$, which is typical for microwave systems \cite{Jin}\cite{Oxborrow}. Note, that due to the high number of accessible RE states, one can manipulate the external electric field to induce the Stark shift and easily match the desired transition wavelength to the crystal length, obtaining a wide range of $\lambda = 2l, l, \frac{2}{3}l ...$ in the same system. This makes our proposal highly tunable.

\begin{figure}[ht!]
\centering
\includegraphics[width=.8\linewidth]{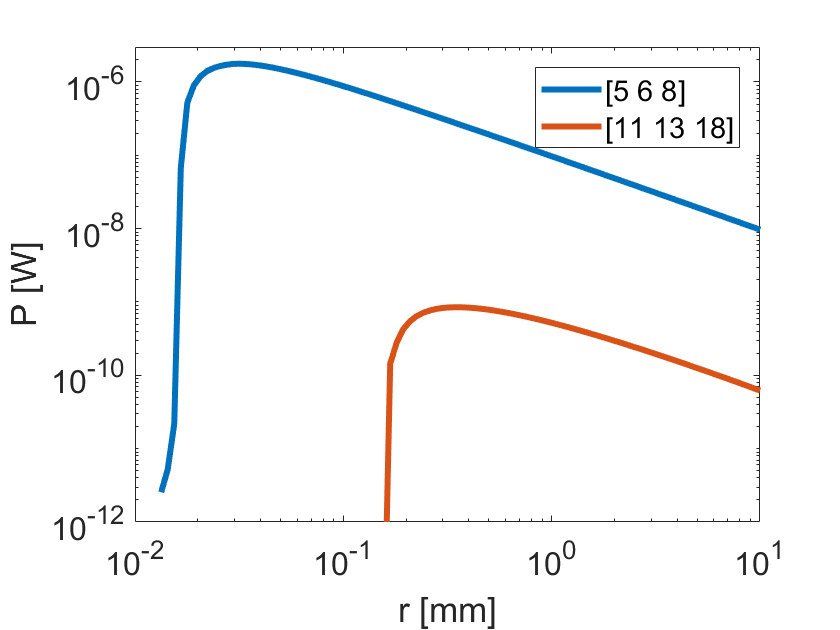}
\caption{Emission power as a function of crystal radius.}\label{rys_R}
\end{figure}

The pump irradiates the crystal from sides, with power density of $\rho_p=$1 mW/mm$^2$ to avoid excessive quenching of high n resonances \cite{Kazimierczuk}.
 For the sake of illustration, we show on the Fig. \ref{rys_Typ} the emission power as a function of time for the above mentioned parameters and for transitions between $n_3=8\rightarrow n_2=6\rightarrow n_1=5$ states which provide efficient maser action.
One can see that the system quickly reaches a steady state with emission power $P \sim 10^{-7}$ W. Other examined configurations are characterized by similar dynamical properties; for higher n states, the initial population oscillations are slowly damped due to the characteristically long lifetimes of these states. To maximize the output power, the important issue is the choice the optimal crystal geometry. Calculations were performed for two particular RE state configurations and a range of values of radius $r$. The results are shown on the Fig. \ref{rys_R}.
A small radius is beneficial due to the high surface to volume ratio; the absorbed pump power is proportional to the side area, while the maximum number of RE depends on crystal volume. For very small $r$, the number of excitons which can be created within considered volume is limited due to the Rydberg blockade. This is particularly important for the considered $18\rightarrow 13\rightarrow 11$ system, which cannot sustain masing below $r \approx 0.15$ mm. To sum up, there is an interplay between the size of the upper level excitons, crystal geometry and the number of excitons. Basing on these results, we have chosen $r = 0.2$ mm to include the higher $n$ states and ensure that the systems based on lower states operate well below Rydberg blockade limit.

\begin{figure}[ht!]
\centering
\includegraphics[width=.9\linewidth]{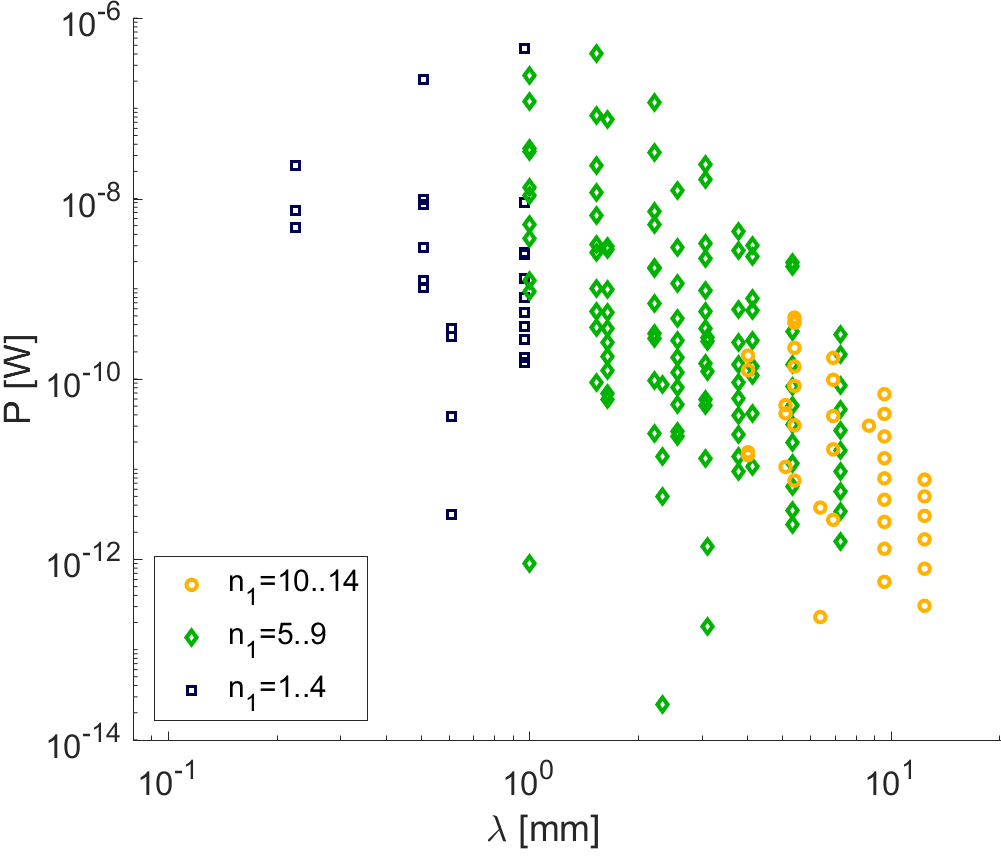}
\caption{Power spectrum of available emission wavelengths for $n_1<n_2<n_3<25$.}\label{rys_spektrum}
\end{figure}

With above mentioned parameters, the emission power has been calculated for all level configurations. The results are shown on the Fig. \ref{rys_spektrum}, where emission power $P$ is presented as a function of wavelength. For $10^{-14}<P<10^{-6}$, a number of wavelengths in the sub-milimeter to centimeter range is available. Results are divided into three groups depending on $n_1$, which has the biggest impact on the emission wavelength. One can see that there are only few systems with very low $n_1<5$. This is caused by the short lifetime of these states, demanding significant pump power to maintain population inversion. On the other hand, the efficiency of high $n$ systems is limited by low exciton populations due to the Rydberg blockade and lower photon energy for these transitions. In all cases where masing has been achieved, $|n_2-n_1|<4$, which is a result of high dipole matrix element for these transitions, ensuring efficient coupling to the cavity. Another important factor is the minimal value of $Q$ needed to obtain sufficient Purcell factor to make $n_3$ state short-lived compared to $n_2$. Therefore, higher quality cavity can support a wider range of state combinations. The results with $P \sim 10^{-12}$ W, visible on the Fig. \ref{rys_spektrum} below the main group, correspond to the dynamic systems where some stimulated emission is present, but no stable inversion is maintained.

\begin{figure}[ht!]
\centering
\includegraphics[width=.85\linewidth]{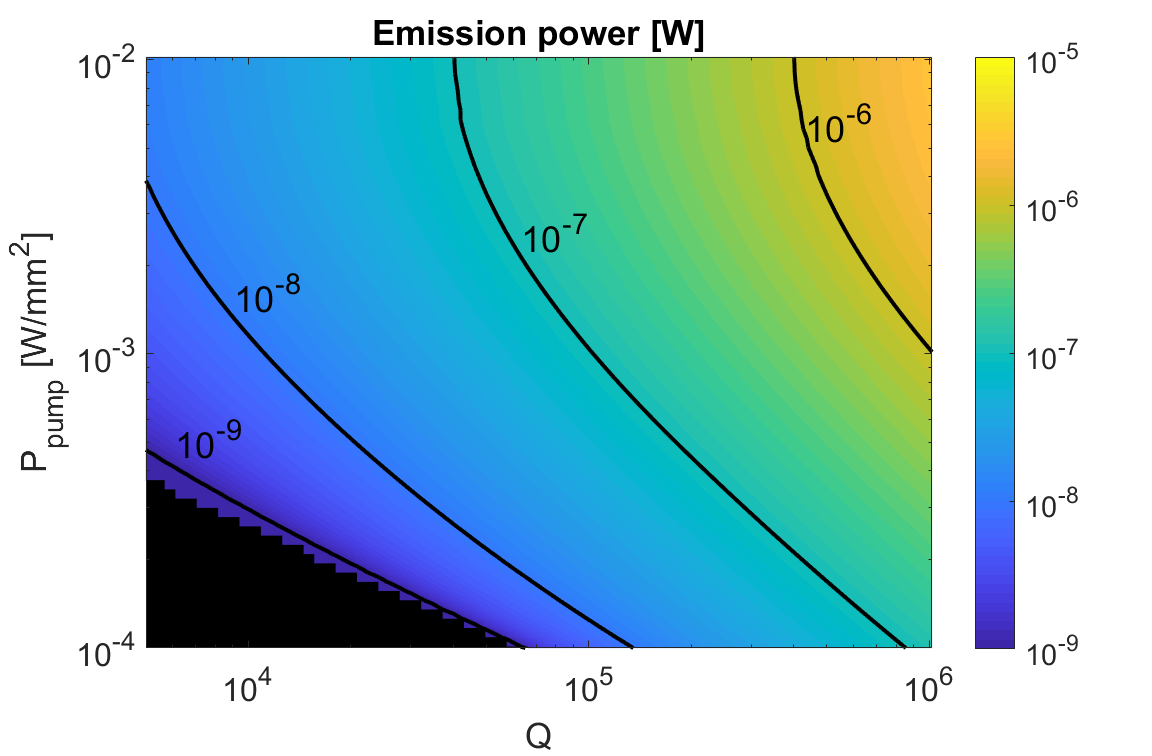}
\includegraphics[width=.85\linewidth]{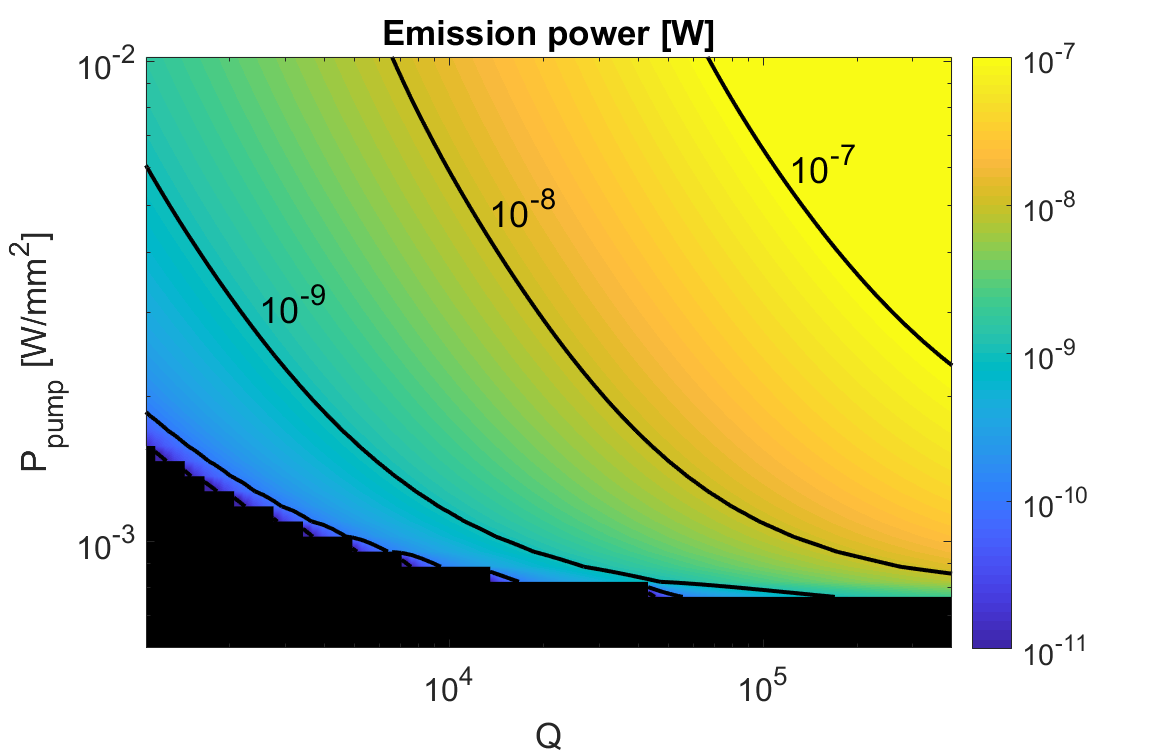}
\caption{Emission power as a function of pump power and $Q$ factor for states \hbox{a) $8\rightarrow6\rightarrow5$} \hbox{and b) $18\rightarrow13\rightarrow11$}}\label{rys_PQ}
\end{figure}

Finally, the dependence of the power on the pump rate and cavity Q factor is shown on the Fig. \ref{rys_PQ}. One can see that the minimal values of these parameters needed for sustaining population inversion and masing action depend on the chosen system. The emission power increases with pump power $P$ and cavity factor $Q$ and is higher for low $n$ configurations. Importantly, even very low $Q$ factor is sufficient to start the masing action. 
\section{Conclusions}
We have demonstrated a proposal for obtaining masing action in RE Cu$_2$O system. The device can be tuned by external electric field and due to the high density of available energy levels, a wide range of wavelengths can be generated. The proposed system is robust and operates in a wide range of pumping power and cavity Q factor.

\section*{Funding}
Support from the National Science Centre, Poland (project
OPUS 2017/25/B/ST3/00817) is greatly acknowledged.
\setcounter{equation}{0}
\renewcommand\theequation{A.\arabic{equation}}
\section*{Appendix A}
\appendix
The Rydberg excitons are modeled by a hydrogen-like wavefunction
\begin{equation}
\psi(r,\theta,\phi)=R(r)Y_{lm}(\theta,\phi),
\end{equation}
where $R(r)$ is the radial part and $Y_{lm}$ are spherical harmonics. The dipole moment is given by
\begin{equation}
\langle \psi_f|er|\psi_i\rangle = \int \psi_f^*er\psi_i d^3r.
\end{equation}
For the $S\rightarrow P$ transitions, $\Delta l=1$ and $m=0$, the angular part of the above equation is a constant \cite{Ditz}
$\int Y_{l_fm_f}^*Y_{l_im_i} \sin \theta d\theta d\phi=\frac{1}{\sqrt{3}}$
while the radial part 
\begin{equation}
R_{if} = 4\pi \int R_{n_f,l_f}^*(r)R_{n_i,l_i}(r) r^3 dr
\end{equation}
is calculated numerically and strongly depends on the principal quantum number. To take the quantum defect into account, one has to use $n^* = n - \delta$.



\end{document}